# Bifurcation enhances temporal information encoding in the olfactory periphery

Kiri Choi,[1, 2, 3, *] Will Rosenbluth,[1, *] Isabella R. Graf,[2, 4, 5] Nirag Kadakia,[1, 2, 3, †] and Thierry Emonet[1, 2, 4, 6, ‡]

[1]*Department of Molecular, Cellular, and Developmental Biology,*
*Yale University, New Haven, Connecticut 06511, USA*
[2]*Quantitative Biology Institute, Yale University, New Haven, Connecticut 06511, USA*
[3]*Swartz Foundation for Theoretical Neuroscience,*
*Yale University, New Haven, Connecticut 06511, USA*
[4]*Department of Physics, Yale University, New Haven, Connecticut 06511, USA*
[5]*Developmental Biology Unit, European Molecular Biology Laboratory, 69117 Heidelberg, Germany*
[6]*Interdepartmental Neuroscience Program, Yale University, New Haven, Connecticut 06511, USA*
(Dated: October 6, 2024)

Living systems continually respond to signals from the surrounding environment. Survival requires that their responses adapt quickly and robustly to the changes in the environment. One particularly challenging example is olfactory navigation in turbulent plumes, where animals experience highly intermittent odor signals while odor concentration varies over many length- and timescales. Here, we show theoretically that *Drosophila* olfactory receptor neurons (ORNs) can exploit proximity to a bifurcation point of their firing dynamics to reliably extract information on the timing and intensity of fluctuations in the odor signal, which have been shown to be critical for odor-guided navigation. Close to the bifurcation, the system is intrinsically invariant to signal variance, and information about the timing, duration, and intensity of odor fluctuations is transferred efficiently. Importantly, we find that proximity to the bifurcation is maintained by mean adaptation alone and therefore does not require any additional feedback mechanism or fine-tuning. Using a biophysical model with calcium-based feedback, we demonstrate that this mechanism can explain the measured adaptation characteristics of *Drosophila* ORNs.

## I. INTRODUCTION

Successful olfactory navigation depends on the ability of animals to make informed navigational decisions, which in turn requires accurate sensory processing to extract relevant information from odor signals. Animals navigate turbulent odor plumes [1–3] by turning upwind when they detect odor and crosswind otherwise [4–13]. For this strategy to be successful, they must extract from the signal not only the intensity of the odor filaments they encounter but also temporal information, such as when and how frequently odor filaments arrive, how long they last, and when they end, see Fig. 1(a) and [4, 8, 14–20]. Information about the timing of odor signals is also critical for detecting the direction of motion of odor filaments [21] and resolving odor sources in space [15, 17]. To reliably extract such information from an odor signal, the olfactory system must adapt its sensitivity to the ambient statistics of the odor concentration, which in turbulent plumes can span orders of magnitude [1–3, 22].

Odor detection starts with odorant molecules binding to olfactory receptors (ORs). Different types of ORs are expressed in the corresponding olfactory receptor neurons (ORNs), which share similar response functions but with shifted sensitivities [23, 24]. In *Drosophila*, ORNs adapt their gain, keeping the steady state firing rate largely invariant in response to changes in mean odor in-tensity (mean adaptation) according to Weber-Fechner's law [25, 26]. Adaptation to the mean signal intensity takes place upstream of the firing machinery at the level of the receptors via a feedback mechanism involving calcium [25–29]. ORNs also intrinsically display finite yet rapid adaptation to stimulus variance, maintaining the variance in the firing rate in response to changes in the stimulus variance via an unknown mechanism that involves both signal transduction and the firing machinery [29] [see Fig. 1(b), inset]. Finally, measurements of the cross-correlations between odor signal and firing rate show that *Drosophila* ORNs maintain a high degree of temporal precision, where the delay between odor onset and firing response stays invariant to mean odor intensity [24, 27, 29, 30]. Together, these observations suggest that gain adaptation to the mean and variance of the odor signal contributes to ORNs' precise encoding of the timing and intensity of fluctuations in odor signals. But the mechanism underlying such precision remains unclear.

Gain control can be mediated through circuit motifs as demonstrated in vision [31–35], olfaction [36], and biochemical systems [37], providing multiple scales of adaptation. Examples from systems neuroscience and molecular signaling show that gain control can also be achieved by more intrinsic and immediate mechanisms [33, 38–44]. For example, the dynamics of receptors and ion channels in a single neuron can induce intrinsic gain scaling [27, 45] when components with different activation timescales are involved [42], and the theory shows that the bifurcation dynamics of a spiking neuron can contribute to contrast gain adaptation [46].

When *Drosophila* ORNs are exposed to a fluctuating

---

* These authors contributed equally to this work
† nirag.kadakia@yale.edu
‡ thierry.emonet@yale.edu

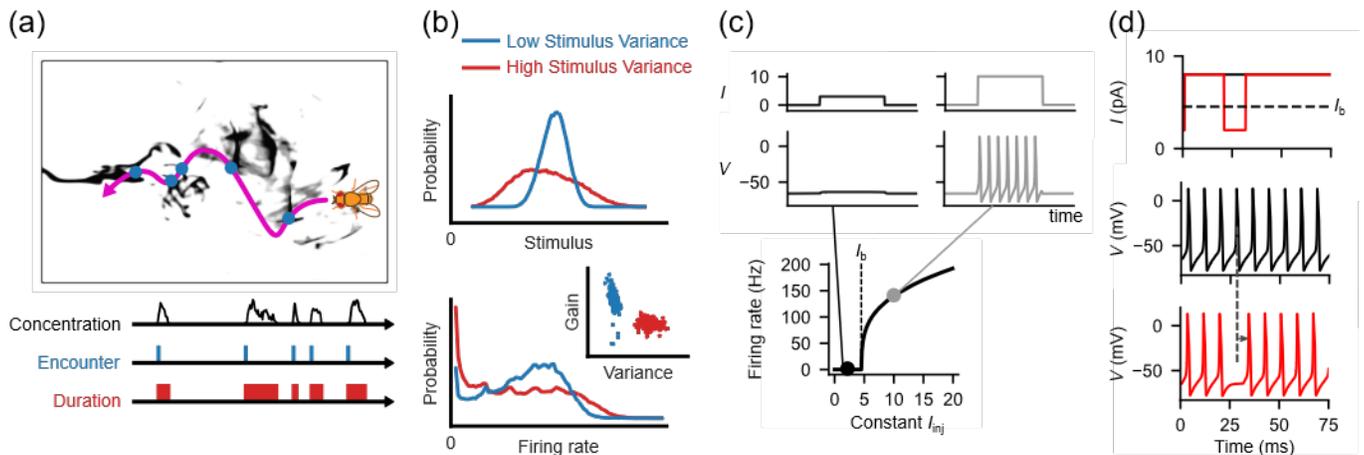

FIG. 1. Encoding the intensity and timing of signal fluctuations. (a) Flies navigate odor plumes using information about the intensity and timing of the odor fluctuations they encounter. (b) The distributions of odor stimulus with low (blue) and high (red) variances (top) and the corresponding ORN firing rate distributions (bottom) [29]. The inset shows changes in ORN gain in response to changes in the stimulus variance. (c) The response dynamics of a neuron spiking above and below the firing threshold. Top: Input current over time for a current temporally above (right, grey) and below (left, black) the firing threshold. Middle: corresponding membrane voltage over time. Bottom: Average firing rate r as a function of a fixed input current. The simulated neuron is quiescent below the threshold current $I_b$, while the firing rate suddenly and monotonically rises with the current above the threshold. (d) For dynamic (non-static) currents, the dose-response curve from (c) may not apply. Top: a static current (black) and a current with a brief dip below the firing threshold (red). Middle and bottom plots: corresponding neuron membrane voltages for two currents. For the current with a dip, the neuron crosses into a non-firing regime and causes a brief delay in spikes, compared to the response to the static current (dotted line and arrow). This delay will manifest as a reduction – but not zeroing – of the firing rate, obscuring the fact that the neuron is crossing into a quiescent state.

odor signal of constant but low mean intensity, the firing rate exhibits a bimodal distribution [29], indicating that the neuron is transitioning between a state of "firing" and "resting" [see Fig. 1(b)]. This observation prompted us to ask whether ORNs use the bifurcation of their firing dynamics to mediate variance adaptation, and if so, the bifurcation might contribute to information encoding. An interesting possibility is that operating near the bifurcation enhances the encoding of the intensity and timing of odor fluctuations, which are critical for navigation. Consistent with this idea, a past study has shown that an *a priori* perfectly gain adapting system that consists of a binary threshold followed by a linear filter can encode more than 1 bit of information when the stimulus exhibits temporal correlations and the signal mean is tuned to the threshold [38].

Here, we investigate if similar encoding properties emerge when neurons operate near a bifurcation of their firing dynamics, and for ORNs, whether this enhances the encoding of the magnitude and timing of odor fluctuations. We address these questions using a Morris-Lecar type model [47], a highly simplified membrane dynamics model in the Hodgkin-Huxley family, whose phase space is fully described in two dimensions. For constant currents below the firing threshold (bifurcation) $I_b$, the neuron does not fire. Above the threshold, the voltage spikes periodically and the firing rate increases monotonically with the injected current [Fig. 1(c)]. But near $I_b$, the distinction between the "firing" and "resting" states is more ambiguous. Consider a brief change in current, from above $I_b$ to below $I_b$ and back above [red line in Fig. 1(d)]. While the neuron has technically switched between two states, this state change is masked in the voltage trace, which only exhibits a delay in spike timing and a brief reduction in firing rate. Thus, for dynamic input currents near $I_b$, the neuron can act as if it is above $I_b$ – encoding signals with a continuous-value firing rate rather than as a discrete two-state ON/OFF system.

We first show that variance adaptation arises intrinsically in the response of individual neurons near their firing threshold, not only for saddle-node on an invariant circle (SNIC) bifurcations [46] but also for Hopf bifurcations. Next, we show that proximity to the bifurcation enhances the encoding of information about the timing, duration, and intensity of input fluctuations, such as those experienced by ORNs in an animal navigating turbulent plumes. In the Supplementary Materials, we also show that these properties can be incorporated into the widely-used linear-nonlinear (LN) model framework by defining the nonlinearity in such model as an approximation of the dynamics near the bifurcation. Finally, we extend the Morris-Lecar model into a full biophysical model of *Drosophila* ORNs incorporating odor-receptor binding and calcium dynamics, which mediate adaptation to the signal mean. By keeping the neuron near the bifurcation, adaptation to the signal mean also provides intrinsic variance adaptation, thereby enhancing the encoding of odor fluctuations over a wide range of signal

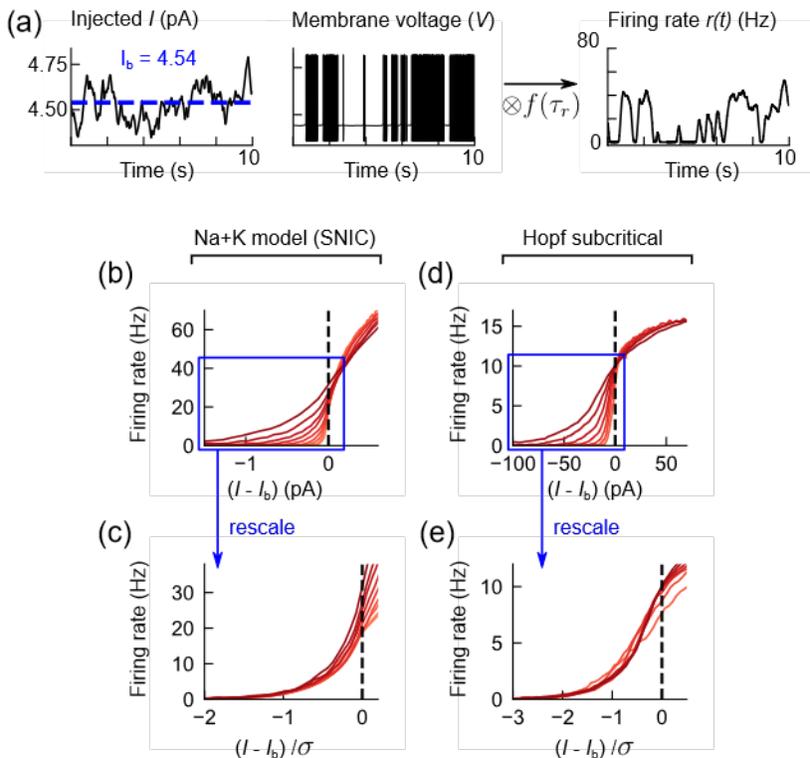

FIG. 2. Dose-response curves of conductance-based neuron models are intrinsically adaptive (invariant) to input signal variance regardless of the bifurcation type. (a) A schematic showing how input current is translated to firing rate in a conductance-based neuron model. Here, the input current $I(t)$ is modeled as an O-U process with input correlation timescale $\tau_s = 200ms$, mean $\mu = 4.54pA$, and standard deviation $\sigma = 0.1$. The firing rates are calculated by binarizing and filtering the spikes with an exponential filter $f$ with firing rate filter timescale $\tau_r$. This model has Na$^+$ and K$^+$ channels driving action potentials and fires when the injected current is above $I_b = 4.54pA$. (b) Dose-response curves of the firing rate as a function of the input current difference from the firing threshold ($I_b = 4.54pA$) obtained from the Na+K model with an SNIC bifurcation. The different curves correspond to increasing fluctuations with size $\sigma$, ranging from $\sigma = 0.04pA$ (light red) to $1.6pA$ (dark red). (c) When rescaled by the magnitude of $\sigma$, the dose-response curves collapse to a single curve, implying that the system inversely adapts its gain with $\sigma$. (d) Dose-response curves from a neuron model with a Hopf subcritical bifurcation, where the firing rate is discontinuous at the threshold current $I_b = 101pA$. (e) The same plot with the x axis rescaled by $\sigma$, as in (c).

intensities.

## II. RESULTS

### A. Intrinsic Variance Adaptation Irrespective of Bifurcation Type

*Drosophila* ORNs intrinsically adapt their response to changes in signal variance [29], allowing them to adjust their sensitivity to the given odor signal. Previous studies have proposed that neurons constantly crossing the firing threshold can implement a simple strategy for intrinsic variance adaptation [38, 46, 48]. Here, we demonstrate intrinsic variance adaption due to proximity to a bifurcation and characterize the adaptive behavior against different bifurcation types. We use a two-dimensional Hodgkin-Huxley conductance-based neuron model known as the Morris-Lecar model (Na+K model) [47], in which inward-flowing Na$^+$ and outward-flowing K$^+$ channels drive spike generation. We drive the neuron model with a fluctuating current $I(t)$ defined by an Ornstein-Uhlenbeck (O-U) stochastic process with input correlation timescale $\tau_s$, standard deviation $\sigma$, and mean $\mu$. Once the spikes are generated, we compute the firing rate $r(t)$ by convolving the binary sequence of threshold voltage crossings with a linear filter with timescale $\tau_r$ [see Fig. 2(a), Appendix A].

When we drive the Na+K model with a fluctuating O-U input current $I(t)$ but with its mean at the bifurcation point (e.g., $\mu = I_b = 4.54pA$), we see a smooth, continuous response much like in a traditional $I$-$r$ curve [see Fig. 2(b)], unlike the dose-response curve derived for constant, non-fluctuating signals [Fig. 1(c)]. However, the $I$-$r$ curve has a distinct dependence on the amount of signal fluctuation $\sigma$. Inputs with higher variance [Fig. 2(b), dark red] have broader $I$-$r$ curves. When the input drops below $I_b$, changes in the firing rate only depend on the excursion time below $I_b$ but not on the magnitude of the drop. Since this excursion time depends on the input cor-relation timescale of the O-U input, rescaling the input by its $\sigma$ collapses the distinct $I$-$r$ curves onto a single curve when $I < I_b$ [see Fig. 2(c)], indicating intrinsic adaption to the variance of the input.

The bifurcation-induced variance adaptation is not unique to this model. Though neurons can differ widely in their response properties, the nature of the quiescence-to-firing bifurcation falls into only a few universality classes [49]. The Na+K model exhibits a saddle node on an invariant circle (SNIC) bifurcation, in which the spiking frequency goes arbitrarily close to zero near $I_b$. Another common bifurcation observed in neurons is a Hopf bifurcation, which exhibits a discontinuity in the frequency-current response and can be either subcritical or supercritical. When modeling a spiking neuron model exhibiting a Hopf bifurcation (see Appendix A), we observed a similar collapse of the $I$-$r$ curves despite the different spiking behavior near the critical current $I_b$ exhibited by the two models [see Figs. 2(d), (e)]. Note that precise adaptation breaks down above the bifurcation point due to the nonlinear yet monotonic relationship between the input and the firing rate in this regime. In a simpler threshold model, where the magnitude of the signal is fully binarized, adaptation occurs both above and below the threshold [38], unlike in the neuron models considered here.

## B. Bifurcation Helps Maintain Coding Capacity Across Signal Variance

When navigating an odor environment, flies need to detect and integrate various features from the odor signal. We quantify the coding capacity of a neuron by calculating the mutual information (MI) between input current $I$ and firing rate $r$, defined as

$$MI(I,r) = \sum_I P(I) \sum_r P(r|I) log P(r|I) - \sum_r P(r) log P(r), \qquad (1)$$

where $P(I)$ and $P(r)$ are the probability distributions of the current and firing rate, respectively (see Appendix A). Like above, we use as input a correlated O-U current with a mean equal to the bifurcation current $I_b$ and verified that the numerical MI calculation is robust to the binning process (see Supplemental Material Sec. 1, Supplemental Material Fig. S1 [50]). Since scaling of the $I$-$r$ curve with signal strength $\sigma$ only occurs when the current is below the bifurcation point [Figs. 2(c), (e)], we consider two separate quantities, MI and MI-, defined as the mutual information calculated over the entire signal and the mutual information calculated over only the times when the signal is below the threshold ($I < I_b$), respectively. MI- quantifies the contribution of the sub-threshold regime to the information transfer. MI- and MI encode different types of information used by animals during odor-guided navigation. Intuitively, mutual information below the bifurcation point (MI-) is correlated to the time since the last odor encounter. In contrast, mutual information over the entire signal (MI) encodes the relative changes in the odor concentration.

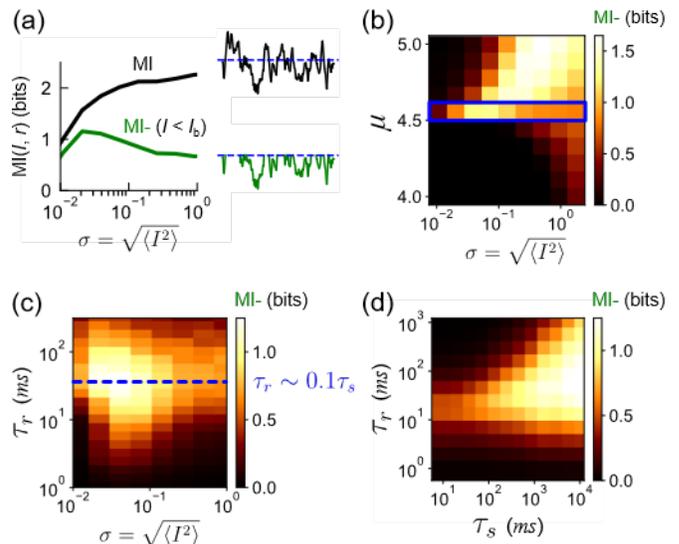

FIG. 3. Bifurcation-induced variance adaptation maintains coding capacity. (a) Mutual information between the distribution of input current $I$ (modeled as an O-U process with mean $\mu$, standard deviation $\sigma$, and correlation timescale $\tau_s s = 500ms$) and the firing rate $r$ as a function of the strength of the fluctuations in the input current (MI, black). Green: same but only for currents below the threshold ($I < I_b$, MI-). MI increases monotonically with the signal variance. Meanwhile, MI- remains largely independent of the signal variance over two orders of magnitude. The firing rate filter timescale is $\tau_r = 55ms$. (b) MI- as a function of mean $\mu$ and standard deviation $\sigma$ of the input current. The blue box corresponds to $\mu \sim I_b = 4.54pA$. (c) MI- as a function of $\sigma$ and $\tau_r$. MI- is maximized when $\tau_r = 40ms$, which is $\sim 1/10$ of the input $\tau_s$ (blue dashed line). (d) MI- as a function of $\tau_s$ and $\tau_r$ shows that information transfer is bounded by the constraint $\tau_r \lesssim 1/5\tau_s$ and large enough input timescale $\tau_s$ to cover the inter-spike interval. Note the change in color scales for (b-d).

We find that MI monotonically increases as $\sigma$ increases over two orders of magnitude [see Fig. 3(a)]. This is expected since a highly variable odor landscape leads to a wider range of signal and response magnitudes. In contrast, MI- is largely independent of $\sigma$ [see Fig. 3(b), Supplemental Material Fig. S2 [50]] since for inputs below $I_b$, adaptation to variance in the input automatically takes place. Furthermore, the MI- distribution is asymmetric with respect to $I_b$, as the amount of encoded information quickly diminishes when $\mu < I_b$. Generally, when $\sigma$ is small and $\mu$ is far from the bifurcation point, the rate of threshold crossing goes down, leading to a decrease in MI. For MI-, this effect is exacerbated when $\mu < I_b$ since spiking occurs too sporadically to maintain the firing rate, eventually yielding zero firing rate for most times when $I < I_b$. Finally, MI- stays elevated

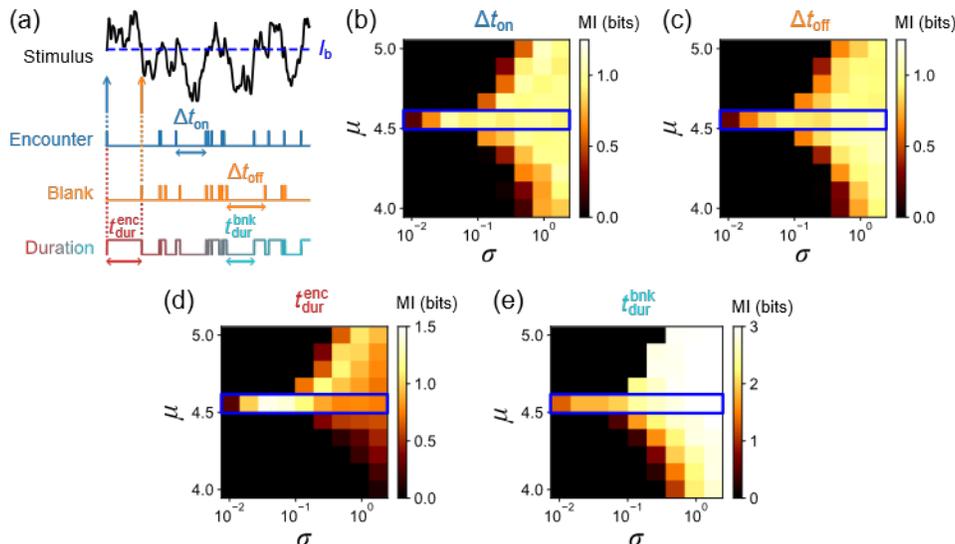

FIG. 4. The coding capacity of signal timing is elevated when ORN is proximal to the bifurcation point. (a) A schematic illustrating different temporal statistics that ORNs may encode for odor-guided navigation. Odor encounters and odor blanks correspond to the stimulus crossing the firing threshold from below/above. The timings between subsequent encounters or blanks are defined as $\Delta t_{ON}$ and $\Delta t_{OFF}$, respectively. The duration of odor encounters $t_{dur}^{enc}$ and blanks $t_{dur}^{bnk}$ denote the time from an encounter to a blank or a blank to an encounter, respectively. (b) MI between the distribution of encounter times $\Delta t_{ON}$ and the firing rate $r$ as a function of signal mean $\mu$ and standard deviation $\sigma$. MI increases and saturates as encounter frequency increases, which depends on the relative values of $\mu$ and $\sigma$. When $\mu$ is close to $I_b = 4.54 pA$ (blue box), information about $\Delta t_{ON}$ is encoded over the widest range of $\sigma$. (c-e) Same as (b) for the distributions of blank times $\Delta t_{OFF}$, encounter duration $t_{dur}^{enc}$, and blank duration $t_{dur}^{bnk}$. Unlike other temporal cues, information about encounter duration becomes small when $\mu < I_b$. Note the changes in color scales for different temporal statistics.

over a range of filter timescales $\tau_r$ [Figs. 3(c), (d)].

MI characteristics are closely related to the mechanistic origin of the variance adaptation near $I_b$. When the model neuron generates spike events from an input, this amounts to a binarization of the signal [Fig. 2(a)]. If we considered the mutual information between a static, temporally uncorrelated signal and its response [as in Fig. 1(c)], MI would not exceed 1 bit and MI- would be 0. However, in the neuron model, the observed MI is significantly higher than 1 bit due to the temporal correlation present in the input signal and thereby in the firing rates [38, 51]. Since this correlation persists over the input correlation timescale $\tau_s$, the coding capacity of the neuron depends on the relative magnitude of the input timescale $\tau_s$ and the filtering rate timescale $\tau_r$. When $\tau_r \gg \tau_s$, all fluctuations average out, and MI barely exceeds 0 bit [see Fig. 3(d)]. When $\tau_r \ll \tau_s$, the output converges to a binary time series, and MI does not exceed 1 bit (and 0 bit for MI-). Calculating MI- for a range of input and filtering timescales, we find that the optimal coding capacity lies on the manifold $\tau_r \sim \tau_s/10$ [Fig. 3(d), Supplemental Material Fig. S3 [50]], consistent with observations in a simpler binary switching model [38].

This insight can be exploited to improve coding capacity in traditional LN neuron models (see Supplemental Material Sec. 2 [50]) consisting of a Gaussian filter followed by a rectified linear unit (ReLU). Standard LN models do not exhibit variance adaptation (Supplemen-tal Material Fig. S4, red [50]). However, substituting the ReLU for the SNIC bifurcating $I$-$r$ curve [as in Fig. 1(c)], we see an imperfect gain control (Fig. S4, orange [50]) since the response changes are magnified for transitions between the active and quiescent states but diminished within the active state. Additionally, switching the order of operation from LN to nonlinear linear (NL) shows a significant improvement ($\sim 15\%$) in coding capacity [see Fig. S4(c) [50]] while keeping the gain adaptation. Compared to standard LN models, this new architecture better reflects the generation of firing rates from spiking neurons and significantly improves the approximation of sensory neuron dynamics.

## C. Bifurcation Enhances Information Encoding from Temporal Odor Cues

In turbulent odor environments, animals use the timing of their encounters with odor filaments to decide when to orient upwind or crosswind [2, 8, 20, 52, 53]. Here, we examine how the ORN's ability to encode odor timing information varies with the proximity to the bifurcation. We used the critical current value $I_b$ as the threshold to define when the signal is detected or not. As temporal cues of interest, we consider $\Delta t_{ON}$, the time difference between subsequent positive crossings of the threshold; $\Delta t_{OFF}$, the time difference between subsequent neg-ative crossings of the firing threshold; $t_{dur}^{enc}$, the duration of "odor encounters" (signal above the firing threshold); and $t_{dur}^{bnk}$, the duration of "blanks" (signal below the firing threshold) [see Fig. 4(a)]. From our simulations, we obtain the stationary distributions of $\Delta t_{ON}$, $\Delta t_{OFF}$, $t_{dur}^{enc}$, and $t_{dur}^{bnk}$ and calculate the mutual information between these distributions and that of the firing rate $r$.

We find that significant information about the signal timing is encoded across a range of mean $\mu$ and standard deviation $\sigma$ of the input current [see Figs. 4(b)-(e)]. Furthermore, when $\mu \sim I_b$, information about the signal timing stays elevated over more than two orders of magnitude of $\sigma$ [blue box in Figs. 4(b)-(e)]. The MI distributions for $\Delta t_{ON}$ and $\Delta t_{OFF}$ are similar and symmetric along the firing threshold $I_b$. Interestingly, MI from encounter duration $t_{dur}^{enc}$ lacks symmetry along $I_b$, as the amount of encoded information diminishes when $\mu < I_b$ [Fig. 4(d)]. There is a general trend of increasing MI as the input variance increases, reflecting the fact that a higher frequency of threshold crossing translates to an increase in coding capacity. However, MI saturates as the threshold crossing frequency saturates, which is limited by the timescales of the O-U process.

Encounter frequency and duration are critical components of the *Drosophila* odor-guided navigation strategy in turbulent plumes [8, 11, 21]. Interestingly, the ORN efficiently transmits information about the duration of odor encounters $t_{dur}^{enc}$, but only when $\mu \geq I_b$ [Fig. 4(d)], suggesting that odor encounter duration is less informative when the signal intensity is low, which is typically the case in turbulent plumes [8, 11, 16, 18]. The MI of the $t_{dur}^{enc}$ depends asymmetrically on $\mu - I_b$ [see Fig. 4(d)] because the duration of an encounter effectively disregards information encoded by fluctuations that take place when the signal is above $I_b$. When $\mu < I_b$, any fluctuations above $I_b$ result in a short encounter duration, which carries no information about the magnitude of the fluctuation. However, when $\mu > I_b$, the signal below $I_b$ rapidly drives the firing rate down to the lower bound of 0 $Hz$. This also explains why the blank duration does not exhibit asymmetry along $\mu \sim I_b$, as blank duration only tracks signals below $I_b$. We can test this effect by modifying the O-U signal such that the O-U signal above $I_b$ is discretized as a single value, thereby creating an upper bound for the firing rate. Indeed, with this modified O-U signal, the MI of the $t_{dur}^{enc}$ becomes symmetric around $I_b$ (see Supplemental Material Fig. S5 [50]).

Our results are robust to different types of input signals. We calculated the mutual information between odor timing and firing rate using as input experimentally measured time traces of odor signals experienced by walking flies freely navigating complex odor plumes [8]. In these traces, the distributions of odor encounter durations and blank durations exhibit near power-law behavior over a few decades [see Fig. 1(i) in [8]] and therefore are very different from the Gaussian statistics of an O-U signal. Nonetheless, we found that the MI distributions are qualitatively similar to those obtained in response to the O-U process [compare Supplemental Material Fig. S6(b) [50] to Fig. 4]. We also check that the MI becomes smaller when using temporally decorrelated Gaussian white noise as input, as expected (Fig. S5 [50]).

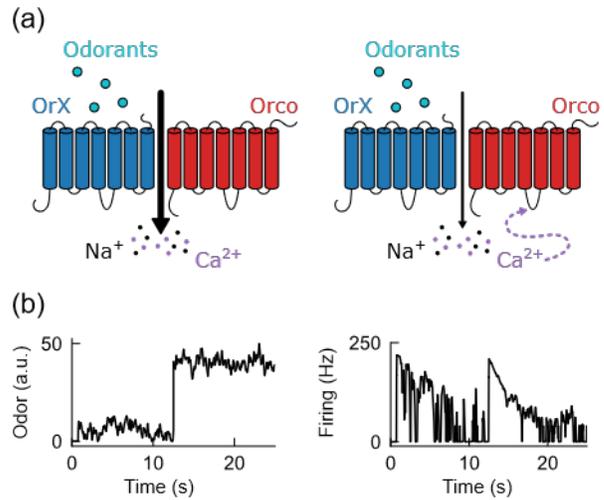

FIG. 5. Overview of the biophysical model of ORNs with calcium-mediated mean adaptation. (a) Mean adaptation in *Drosophila* ORNs is mediated by cytoplasmic calcium level. Odorant drives response by binding to olfactory receptors (OrX). Active receptors drive calcium influx along with sodium ions, which, in return, regulates the receptor activity through a calmodulin-based process or calcium-dependent phosphatase-based dephosphorylation that targets Orco, decreasing the receptor activity. Meanwhile, cytoplasmic calcium ions are sequestered in the mitochondria or pumped out, slowly recovering the sensitivity towards a fixed steady state under a continuous odor input. (b) The firing rate adapts to about $\sim 30\ Hz$ independent of the background odor concentration, as shown experimentally in ORN ab3A in response to ethyl acetate. To leverage bifurcation-induced benefits, we chose a steady state firing rate that corresponds to a signal slightly above the bifurcation point. Left: an odor signal trace where the mean concentration changes by an order of magnitude from 4.5 to 45 A.U. but with the same variance. Right: while the increase in odorant concentration creates a transient increase in firing rate, the ORN quickly adapts to the same average firing rate of $\sim 30\ Hz$.

### D. Mean Adaptation Maintains ORNs Near the Bifurcation Point

So far, we have shown that proximity to a bifurcation enhances the transmission of olfactory information through a binary switching process. In practice, this requires that the signal mean is always positioned close to the bifurcation point $I_b$, which is an ostensibly stringent condition. How do ORNs maintain encoding fidelity in naturalistic conditions, where background stimuli intensities vary substantially between environments?

Various sensory neurons exhibit adaptation to the signal mean [54–58]. A simple mechanism of mean adapta-tion is negative integral feedback, in which the cumulative deviation of a system's response from the baseline is fed back as an inhibitory input, pushing the system back as the response deviates from the baseline. Such negative feedback mechanisms have been observed in cellular [59, 60] and neural systems, including *Drosophila* olfaction [25, 27, 29]. The adaptation in a sensory neuron is often associated with receptors and slow ion channel dynamics [61]. In particular, the cytoplasmic calcium level is a critical component of adaptation in *Drosophila* ORNs [25, 29, 62, 63], suggesting the calcium influx through *Drosophila* OR is a good candidate to model the adaptive behavior.

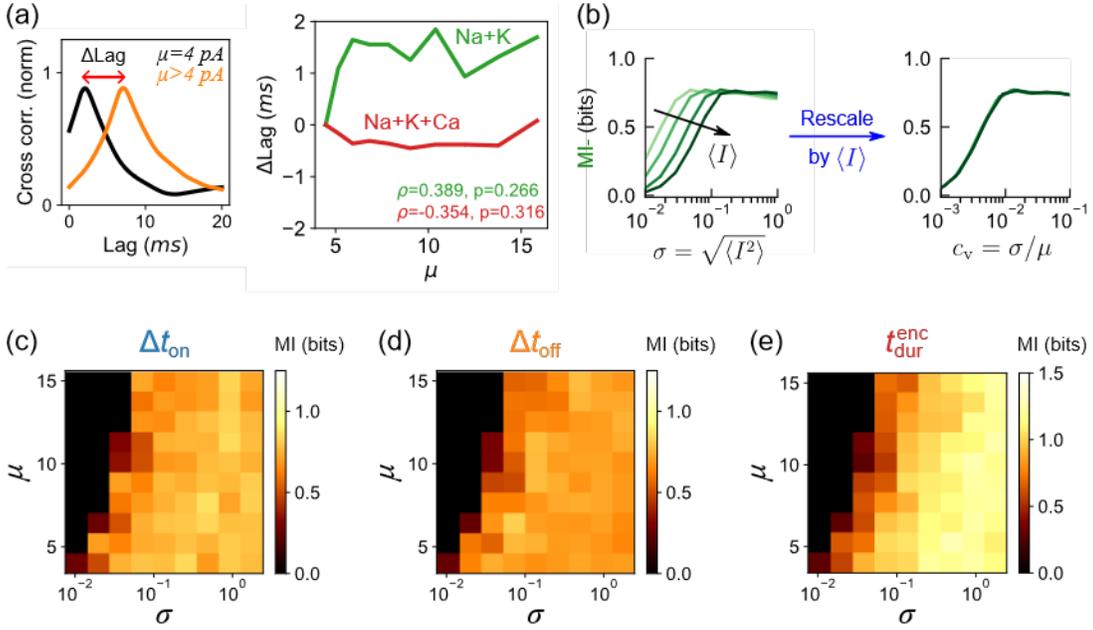

FIG. 6. Mean adaptation maintains robust bifurcation-induced information transmission across stimulus backgrounds. (a) Left: the lag between the input signal and the firing rate is quantified from the peak in the cross-correlation between the two time traces. Right: differences in firing lag ($\Delta$Lag) with respect to the lag at $\mu = 4pA$ (right, black line) as a function of the mean signal intensity $\mu$ for Na+K (green) and Na+K+Ca (red) models. The input signal standard deviation is $\sigma = 0.4$. The firing lag does not increase significantly with the mean signal intensity. $\rho$ and $p$ denote the Spearman correlation coefficient and the corresponding $p$-value. (b) MI- from the Na+K+Ca model using O-U odor signals with $\mu$ ranging from 4 to 15 A.U. and $\sigma$ ranging from 0.01 to 3.16. Once rescaled by the signal mean, the curves collapse (right). (c-e) MI between the firing rate and the distributions of encounter times $\Delta t_{\text{ON}}$, blank times $\Delta t_{\text{OFF}}$, and encounter duration $t_{\text{dur}}^{\text{enc}}$, as a function of $\mu$ and $\sigma$ given the firing rate threshold of $\sim 30\ Hz$.

We extend the conductance-based model (Na+K model) used above into a biophysical model of a *Drosophila* ORN (Na+K+Ca model) by incorporating the calcium influx through OR and cytoplasmic calcium-mediated regulation of the receptor activity [64] (see Appendix A). We model adaptation assuming that the cytoplasmic calcium dynamics are determined by the influx and the outflux of calcium ions based on the activation of the OR and the calcium pump,

$$\frac{dC}{dt} = \frac{1}{\tau_c}(g_s S - g_c C). \qquad (2)$$

Here, $C$ is the cytoplasmic calcium concentration [$Ca^{2+}$], $S$ is the odor concentration, $\tau_c$ is the adaptation timescale, and $g_s$ and $g_c$ are constants [Fig. 5(a)]. The ORs, together with the coreceptor Orco, form non-selective cation channels permeating sodium and calcium ions [63], and previous studies have shown that the adaptation behavior is independent of the firing activity and results from odorant binding to ORs [27, 29]. We assume an active inhibition of OR by calcium ions, perhaps through a calmodulin-based mechanism [65] or calcium-dependent phosphatase-based dephosphorylation that targets Orco [62, 66] [Fig. 5(b)]. Assuming odor binding and unbinding is fast, we model the dependency of the input current $I_s$ on odor concentration $S$ and calcium concentration $C$ as

$$I_s = \frac{g_I S}{K_s + S + C/K_c}, \qquad (3)$$

where $g_I$ is a proportionality constant, $K_s$ controls the sensitivity to the odor, and $K_c$ determines the strength of the inhibition via calcium. Thus, the model contains a negative feedback loop modulated by calcium concentration, which relies on the fast sodium influx and the slow calcium extrusion. Combining Eqs. 2 and 3 with the conductance-based model (see Appendix A), our biophysical model responds to fast changes in signal amplitude and adapts back to $\sim 30\ Hz$ following a step increase in odor concentration in accordance with experimental data [29, 67] irrespective of the background odor level [see Fig. 5(b)].

The biophysical model also reproduces several key dynamical properties of ORNs. First, the firing lag, defined as the peak of the cross-correlation between the input and the response, is nearly invariant to the changes in the signal mean $\mu$ [see Fig. 6(a)] as measured in ORNs [29]. This implies that our model remains temporally precise over diverse odor concentrations. The absolute lag ($\sim 7$ ms) suggested by the model is comparable to the smallest value of experimentally reported odor-evoked first spike latency in *Drosophila* [68]. We also confirmed that the biophysical model reproduces the gain adaptation properties of ORNs, namely the gain is inversely proportional to the mean odor intensity, thus following Weber-Fechner's law [25, 29, 67] (Supplemental Material Fig. S7 [50]), and scales with the signal variance qualitatively similarly to the experiments (Supplemental Material Fig. S8 [50]) [29].

To quantify the coding capacity of the biophysical model, we stimulated it with odor signals modeled as O-U processes. As before, MI- is elevated over a range of signal fluctuations. Moreover, because the gain scales inversely with the mean signal intensity, plotting MI- versus $\sigma/\mu$ collapses all plots onto a single curve [see Fig. 6(b)]. Therefore, the system transmits information about the fluctuations in a manner that is largely invariant with respect to the mean signal intensity. To analyze how information about odor timing is transmitted, we defined odor encounters and blanks using the baseline firing frequency of 30 $Hz$ as the crossing point instead of $I_b = 4.54pA$ since mean adaptation dynamically adjusts the firing threshold. This can be viewed as a "novelty detection," where sudden deviations from the steady state are identified. We observe that adaptation to the mean signal intensity enables the model to consistently encode the timing and intensity of odor fluctuations over a wide range of mean odor concentration and variance [see Figs. 6(c)-(e)].

These results demonstrate that by maintaining their operation in the vicinity of the bifurcation via a calcium-mediated adaptation mechanism, ORNs can robustly encode the intensity and timing of fluctuations in odor signal, which are critical for odor-guided navigation.

## III. DISCUSSION

### A. Proximity to a Bifurcation Enhances Information Transmission in a Model Neuron

In this work, we quantify information transmission in a model ORN via the mutual information between the firing rate and summary statistics of the signal (odor) input. Previous experimental data has shown that when stimulated with a fluctuating odor signal of constant mean, the steady-state firing rate distribution of an ORN tends to be bimodal with a peak at zero firing rate, indicating transitions of the neuron between the firing and resting regimes [29]. Here, we show that proximity to the bifurcation is beneficial for olfactory information processing by providing robust encoding of information about the timing and intensity of fluctuations in odor signals. We also show that proximity to the bifurcation entails intrinsic variance adaptation, as is the case in many other sensory systems [69, 70]. Finally, we show that there is synergy between mean and variance adaptation: ORN adaptation to the signal mean, which follows Weber's law, naturally drives the ORN dynamics to the bifurcation and thereby maximizes information transmission without the need for an additional tuning mechanism.

Note that the firing rate can simultaneously encode different types of information when neurons are near their bifurcation point; When the neuron is below the bifurcation point, the firing rate only encodes information about the recent bifurcation threshold crossing. Above the bifurcation point, the firing rate also encodes information about the stimulus intensity. When the neuron crosses the bifurcation point, information about temporal cues are encoded. Therefore, proximity to a bifurcation is crucial for effectively encoding temporal cues and extracting more information from temporally correlated stimuli. Other computational neuron models, such as the LN model, can also benefit from these properties when approximated dynamics near the bifurcation are implemented as the nonlinearity in the model (Supplemental Material Sec. 2 [50]).

Even though the dose-response curves are smooth [see Fig. 2], the bifurcation point remains crucial for information coding capacity. When inputs are noisy or neurons adapt, dose-response curves, which reflect average firing rates, exhibit a finite slope everywhere [71, 72]. However, the slope is steepest near the bifurcation point, where ORNs are most responsive to signal fluctuations. This allows neurons to effectively capture a wide range of signal variations, improving their ability to encode information in the distribution of instantaneous firing rates $r$.

### B. Intrinsic Gain Control in *Drosophila* ORNs

Adaptation to variance in the input stimulus is observed in all sensory modalities [29, 73–78], and several intrinsic mechanisms for gain control have been suggested [33, 38–42]. Here, we showed that not only a SNIC bifurcation but also a Hopf bifurcation can mediates variance adaptation suggesting that this mechanism could be a general mechanism for effective information processing at the single-neuron level.

In our biophysical model of *Drosophila* ORNs, variance adaptation occurs intrinsically as a consequence of mean or spike frequency adaptation via calcium. Interestingly, due to the large susceptibility at the bifurcation point, an adapted firing rate anywhere in the steep part of the response curve will keep the neuron very close to the firing-quiescent bifurcation. For instance, if the firing rate of *Drosophila* ORNs adapts to approximately 30 $Hz$ as we have shown previously in vivo [29], the corresponding injected current is within 0.5 $pA$ of the bifurcation point of our model [see Fig. 1(c)]. ORNs may therefore be able to retain a state of criticality without the need for fine-tuning. This robust tuning to a critical (bifurcation) point with enhanced signal sensitivity is reminiscent of other sensory systems, such as the thermal sensing of pit vipers [79], chemosensing in *E. coli* [80, 81], and sound amplification in the inner ear [69, 82–86]. The underlying feedback motif also belongs to a wider class of integral feedback mechanisms, which exhibit precise adaptation even in networks with noisy dynamics [87].

## C. Implication for olfactory navigation

During olfactory navigation, insects rely heavily on the timing and relative intensity of odor fluctuations to make behavioral decisions [4, 8, 10, 11, 16, 18, 88, 89]. One of the challenges their olfactory system faces during this task is to detect these fluctuations, given that their intensity and timing can vary over orders of magnitude due to turbulence. Mean adaptation allows ORNs to remain sensitive regardless of the mean odor concentration, while variance adaptation mediated by the bifurcation enhances the ability of ORNs to detect the relative intensity and timing of fluctuations. Our model shows that together, these two adaptation mechanisms enhance the coding capacity of information relevant to navigation that is carried by these fluctuations. This is consistent with previous results that show that spiking neurons can adapt while maintaining sensitivity to fluctuations [90].

For example, we find that the frequency and duration of odor encounters are well-encoded by ORNs. We also find that their information content varies depending on context. Odor encounter durations are less informative when odor concentration is low as is typically the case in turbulent plumes. This is consistent with experimental findings showing that walking flies rely on odor duration to navigate diffusion-dominated plumes but less so in complex plumes, where odor encounters are brief and blanks much longer [8, 11, 18]. Finally, we found that proximity to the bifurcation also enhanced information transmission when we used experimentally measured odor signals (instead of an O-U process) as input to our model. Specifically, we used time traces of odor encounters experienced by freely navigating flies in a turbulent plume [8]. The statistics of these odor signals are very different from those of an O-U process. Nonetheless, we found that the MI distributions from the signal timing are similar for both the nonadapting [Na+K model; Fig. S6(b) [50]] and the adapting models [Na+K+Ca model; Fig. S6(c) [50]]. This implies that the large information transmission at the bifurcation applies independently of the plume characteristics and has relevance in real-world odor-guided navigation.

## ACKNOWLEDGMENTS

All original data and codes are deposited and publicly available [91]. We thank Michael Abbott, Sam Brudner, Damon Clark, Viraaj Jayaram, and the members of the Emonet laboratory for helpful discussions and constructive comments on the manuscript. Research reported in this publication was supported by NINDS award RF1NS132840. K.C. was supported by a postdoctoral fellowship through the Swartz Foundation for Theoretical Neuroscience. I.R.G. was supported by the Deutsche Forschungsgemeinschaft Projektnummer 494077061. N.K. was supported by postdoctoral fellowships NIH F32MH118700 and NIH K99DC019397.

K.C., W.R., I.R.G., N.K., and T.E. designed research; K.C., W.R., and N.K. performed research; K.C., W.R., N.K. and T.E. analyzed data; and K.C., W.R., I.R.G., N.K., and T.E. wrote the paper.

## Appendix A: Materials and Methods

### 1. Odor signal modeling

A stochastic naturalistic odor signal environment is simulated as an Ornstein-Uhlenbeck process, defined as

$$S(t + dt) = S(t) - \frac{1}{\tau_s} S(t) dt + c^{\frac{1}{2}} N(t) (dt)^{\frac{1}{2}}, \quad (A1)$$

where $\tau_s$ is the input correlation timescale, $c$ is the diffusion constant, and $N(t)$ denotes an independent normal-distributed random variable at each time t with mean 0 and variance 1. To center the signal around a given mean, we added $\mu$ to $S(t)$. The variance of the odor signal $S(t)$ is equal to $\sigma^2 = c\tau_s/2$.

### 2. Conductance-based neuron model

Previous research demonstrated that adaptation in *Drosophila* ORNs originates from ligand-receptor binding dynamics rather than neuronal firing [27], indicating a conductance-based neuron model is necessary. A simple model satisfying these constraints is a two-dimensional Hodgkin-Huxley type model known as the Morris-Lecar model [47]. This model tracks potassium and sodium ion concentrations in the intra- and extracellular media as well as a leak current and exhibits a SNIC bifurcation. The potential difference across the neuronal cell membrane $V(t)$ and the fraction of potassium channel gates in the open state $n(t)$ are defined by the following sets of differential equations,

$$\frac{dV}{dt} = (I_s + g_L(E_L - V) + g_{Na}m_\infty(V)(E_{Na} - V) + ng_K(E_K - V))/C \tag{A2a}$$

$$\frac{d}{dt}n(t) = \frac{(n_\infty - n)}{\tau_n}, \tag{A2b}$$

where $m_\infty = (1 + e^{\frac{V_m - V}{k_m}})^{-1}$ and $n_\infty = (1 + e^{\frac{V_n - V}{k_n}})^{-1}$.

The parameters are set as $g_L = 8$, $g_{Na} = 20$, $g_K = 10$, $E_L = -80$, $E_{Na} = 60$, $E_K = -90$, $C = 1$, $V_m = -20$, $V_n = -25$, $k_m = 15$, $k_n = 5$, $\tau_n = 1$. An Ornstein-Uhlenbeck signal $S(t)$ is used as the input $I_s$. A timestep of $0.05 ms$ is used to iterate the equations by the Euler method unless noted otherwise. Simulations are run for at least 20 seconds.

In principle, this model contains two voltage-gating variables, $n$ and $m$, where $n$ controls the fraction of open potassium channels, and $m$ controls the fraction of open sodium channels. The characteristic timescale of the gat-ing variable $m(t)$ is assumed to be much faster than fluctuations in $V(t)$ and treated as an instantaneous function of $V$ [$m_\infty(V)$]. Therefore, the current position of the system at any given time can be fully described by the values of $V$ and $n$. The dynamics of other ions present in the intra- and extracellular media, such as calcium and chlorine, are not explicitly considered except when the mean adaptation is implemented.

To test the robustness of variance adaptation across different bifurcation types, we used the Hopf bifurcating model described in [92]. Here, the slow current is driven by calcium, and some parameters are modified to produce a Hopf bifurcation.

$$\frac{dV}{dt} = (I_s + g_L(E_L - V) + g_{Ca}m_\infty(V)(E_{Ca} - V) + wg_K(E_K - V))/C \tag{A3a}$$

$$\frac{dw}{dt} = \lambda_w(w_\infty - w) \tag{A3b}$$

$$\lambda_w = \phi \cdot \cosh\left(\frac{V - V_w}{2k_w}\right) \tag{A3c}$$

$$m_\infty = \frac{1}{2}\left(1 + \tanh\left(\frac{V - V_m}{k_m}\right)\right) \tag{A3d}$$

$$w_\infty = \frac{1}{2}\left(1 + \tanh\left(\frac{V - V_w}{k_w}\right)\right) \tag{A3e}$$

The parameters are defined as: $g_L = 2$, $g_{Ca} = 4$, $g_K = 8$, $E_L = -60$, $E_{Ca} = 120$, $E_K = -84$, $C = 20$, $V_m = -1.2$, $V_w = 2$, $k_m = 18$, $k_w = 30$, $\phi = 0.04$.

### 3. Calculating the firing rate

Hodgkin-Huxley type models describe the evolution of membrane voltage with an inter-spike time on the order of milliseconds, from which we calculate the evolution of the firing rate over hundreds of milliseconds or seconds. To obtain a continuous value firing rate from the voltage-time trace generated by the model, we first run a peak-finding algorithm that binarizes all points along the time trace. The peak-finding algorithm determines a peak based on voltage spikes exceeding a membrane potential of $0\,V$. Then, a convolutional Gaussian smoothing filter with unity time-integral area is applied to the binary data to obtain a time-averaged firing rate. The standard deviation of the filter is given by $\tau_r * dt$, where $\tau_r$ is the firing rate filter timescale and $dt$ is the simulation

step size. Therefore, the filter is defined as

$$g(t) = \frac{1}{(\tau_r dt)\sqrt{2\pi}}e^{-\frac{1}{2}\left(\frac{t}{\tau_r dt}\right)^2}. \tag{A4}$$

### 4. Mutual information

The mutual information between the stimulus and the firing rate is defined as the entropy of the rate minus the entropy of the rate given the corresponding value of the stimulus:

$$MI(rate, stim) = H(rate) - H(rate|stim), \tag{A5}$$

where the entropy is defined by

$$H(rate) = -\sum P(rate)log_2(P(rate)) \tag{A6}$$

and the conditional entropy is defined by

$$H(rate|stim) = \sum P(stim = s) \sum H(rate|stim = s). \tag{A7}$$

Numerically, mutual information can be calculated using the distributions of the stimulus and the (conditional) distributions of the firing rate. We generated a probability density function of firing rate $P(rate)$ using a histogram with 100 equal-sized bins spanning from 0 to 200 $Hz$. The entropy is computed as $H(rate) = -\sum P(rate) * log_2 P(rate) dr$ where $dr$ would be the bin size (=2) when computed discretely. We confirmed that the number of data points is large enough such that the numerical MI values are robust to the error introduced by the binning process (see Supplemental Material Sec. 1, Fig. S1 [50]). To calculate the conditional entropy, we also generated a probability density function for the odor signal by creating a histogram with 100 bins, with an adaptive bin size chosen such that the histogram spans $\pm 3\sigma$ of the input mean $\mu$. When calculating mutual information about temporal odor cues, $3\sigma$ from the sample mean has been used. For each discretized stimulus bin $s$, we collected the distribution of firing rates observed at all corresponding times where $stim = s$, which approximates $P(rate|stim = s)$. By definition, $H(rate|stim = s) = -\sum P(rate|stim = s) * log_2 P(rate|stim = s) dr$ for each stimulus bin $s$, and the conditional entropy $H(rate|stim)$ is equal to the discrete sum of all stimulus bins times the stimulus probability.

### 5. Biophysical *Drosophila* ORN model with adaptation

As shown in the main text, we incorporate calcium-mediated adaptation by expanding upon the previously defined conductance-based neuron model (Eqs. 2 and 3). The parameters in Eqs. 2 and 3 are as follows: $g_s = 0.76875$, $g_c = 0.0625$, $\tau_c = 250$, $g_I = 500$, $K_s = 0.1$, $K_c = 1$. A timestep of 0.1 $ms$ is used for numerical integration.

[1] A. Celani, E. Villermaux, and M. Vergassola, Odor landscapes in turbulent environments, Physical Review X **4**, 041015 (2014).

[2] G. Reddy, V. N. Murthy, and M. Vergassola, Olfactory sensing and navigation in turbulent environments, Annual Review of Condensed Matter Physics **13**, 191 (2022).

[3] J. A. Riffell, L. Abrell, and J. G. Hildebrand, Physical processes and real-time chemical measurement of the insect olfactory environment, Journal of chemical ecology **34**, 837 (2008).

[4] A. Mafra-Neto and R. T. Cardé, Fine-scale structure of pheromone plumes modulates upwind orientation of flying moths, Nature **369**, 142 (1994).

[5] J. Murlis, J. S. Elkinton, and R. T. Carde, Odor plumes and how insects use them, Annual review of entomology **37**, 505 (1992).

[6] C. Flügge, Geruchliche raumorientierung von drosophila melanogaster, Zeitschrift für vergleichende Physiologie **20**, 463 (1934).

[7] J. S. Kennedy and D. Marsh, Pheromone-regulated anemotaxis in flying moths, Science **184**, 999 (1974).

[8] M. Demir, N. Kadakia, H. D. Anderson, D. A. Clark, and T. Emonet, Walking drosophila navigate complex plumes using stochastic decisions biased by the timing of odor encounters, Elife **9**, e57524 (2020).

[9] S. A. Budick and M. H. Dickinson, Free-flight responses of drosophila melanogaster to attractive odors, Journal of experimental biology **209**, 3001 (2006).

[10] F. van Breugel and M. H. Dickinson, Plume-tracking behavior of flying drosophila emerges from a set of distinct sensory-motor reflexes, Current Biology **24**, 274 (2014).

[11] E. Álvarez-Salvado, A. M. Licata, E. G. Connor, M. K. McHugh, B. M. King, N. Stavropoulos, J. D. Victor, J. P. Crimaldi, and K. I. Nagel, Elementary sensory-motor transformations underlying olfactory navigation in walking fruit-flies, Elife **7**, e37815 (2018).

[12] E. Balkovsky and B. I. Shraiman, Olfactory search at high reynolds number, Proceedings of the national academy of sciences **99**, 12589 (2002).

[13] A. M. Matheson, A. J. Lanz, A. M. Medina, A. M. Licata, T. A. Currier, M. H. Syed, and K. I. Nagel, A neural circuit for wind-guided olfactory navigation, Nature Communications **13**, 4613 (2022).

[14] P. Szyszka, T. Emonet, and T. L. Edwards, Extracting spatial information from temporal odor patterns: insights from insects, Current Opinion in Insect Science **59**, 101082 (2023).

[15] A. Sehdev, Y. G. Mohammed, T. Triphan, and P. Szyszka, Olfactory object recognition based on fine-scale stimulus timing in drosophila, Iscience **13**, 113 (2019).

[16] V. Jayaram, N. Kadakia, and T. Emonet, Sensing complementary temporal features of odor signals enhances navigation of diverse turbulent plumes, Elife **11**, e72415 (2022).

[17] G. Raiser, C. G. Galizia, and P. Szyszka, Olfactory receptor neurons are sensitive to stimulus onset asynchrony: Implications for odour source discrimination, Chemical senses , bjae030 (2024).

[18] V. Jayaram, A. Sehdev, N. Kadakia, E. A. Brown, and T. Emonet, Temporal novelty detection and multiple timescale integration drive drosophila orientation dynamics in temporally diverse olfactory environments, PLOS Computational Biology **19**, e1010606 (2023).

[19] S. D. Boie, E. G. Connor, M. McHugh, K. I. Nagel, G. B. Ermentrout, J. P. Crimaldi, and J. D. Victor, Information-theoretic analysis of realistic odor plumes: What cues are useful for determining location?, PLoS computational biology **14**, e1006275 (2018).

[20] I. J. Park, A. M. Hein, Y. V. Bobkov, M. A. Reidenbach, B. W. Ache, and J. C. Principe, Neurally encoding time for olfactory navigation, PLoS computational biology **12**, e1004682 (2016).

[21] N. Kadakia, M. Demir, B. T. Michaelis, B. D. DeAngelis, M. A. Reidenbach, D. A. Clark, and T. Emonet, Odour motion sensing enhances navigation of complex plumes, Nature **611**, 754 (2022).

[22] J. Murlis, M. A. Willis, and R. T. Cardé, Spatial and temporal structures of pheromone plumes in fields and forests, Physiological entomology **25**, 211 (2001).

[23] G. Si, J. K. Kanwal, Y. Hu, C. J. Tabone, J. Baron, M. Berck, G. Vignoud, and A. D. Samuel, Structured odorant response patterns across a complete olfactory receptor neuron population, Neuron **101**, 950 (2019).

[24] C. Martelli, J. R. Carlson, and T. Emonet, Intensity invariant dynamics and odor-specific latencies in olfactory receptor neuron response, Journal of Neuroscience **33**, 6285 (2013).

[25] L.-H. Cao, B.-Y. Jing, D. Yang, X. Zeng, Y. Shen, Y. Tu, and D.-G. Luo, Distinct signaling of drosophila chemoreceptors in olfactory sensory neurons, Proceedings of the National Academy of Sciences **113**, E902 (2016).

[26] J. Cafaro, Multiple sites of adaptation lead to contrast encoding in the drosophila olfactory system, Physiological Reports **4**, e12762 (2016).

[27] K. I. Nagel and R. I. Wilson, Biophysical mechanisms underlying olfactory receptor neuron dynamics, Nature neuroscience **14**, 208 (2011).

[28] K. I. Nagel, E. J. Hong, and R. I. Wilson, Synaptic and circuit mechanisms promoting broadband transmission of olfactory stimulus dynamics, Nature neuroscience **18**, 56 (2015).

[29] S. Gorur-Shandilya, M. Demir, J. Long, D. A. Clark, and T. Emonet, Olfactory receptor neurons use gain control and complementary kinetics to encode intermittent odorant stimuli, Elife **6**, e27670 (2017).

[30] P. Szyszka, R. C. Gerkin, C. G. Galizia, and B. H. Smith, High-speed odor transduction and pulse tracking by insect olfactory receptor neurons, Proceedings of the National Academy of Sciences **111**, 16925 (2014).

[31] S. M. Smirnakis, M. J. Berry, D. K. Warland, W. Bialek, and M. Meister, Adaptation of retinal processing to image contrast and spatial scale, Nature **386**, 69 (1997).

[32] S. A. Baccus and M. Meister, Fast and slow contrast adaptation in retinal circuitry, Neuron **36**, 909 (2002).

[33] A. Borst, V. L. Flanagin, and H. Sompolinsky, Adaptation without parameter change: dynamic gain control in motion detection, Proceedings of the National Academy of Sciences **102**, 6172 (2005).

[34] Y. Cui, Y. V. Wang, S. J. Park, J. B. Demb, and D. A. Butts, Divisive suppression explains high-precision firing and contrast adaptation in retinal ganglion cells, Elife **5**, e19460 (2016).

[35] D. Xing, C.-I. Yeh, J. Gordon, and R. M. Shapley, Cortical brightness adaptation when darkness and brightness produce different dynamical states in the visual cortex, Proceedings of the National Academy of Sciences **111**, 1210 (2014).

[36] A. Barth-Maron, I. D'Alessandro, and R. I. Wilson, Interactions between specialized gain control mechanisms in olfactory processing, Current Biology **33**, 5109 (2023).

[37] L. Qiao, W. Zhao, C. Tang, Q. Nie, and L. Zhang, Network topologies that can achieve dual function of adaptation and noise attenuation, Cell systems **9**, 271 (2019).

[38] I. Nemenman, Gain control in molecular information processing: lessons from neuroscience, Physical biology **9**, 026003 (2012).

[39] S. Hong, B. N. Lundstrom, and A. L. Fairhall, Intrinsic gain modulation and adaptive neural coding, PLoS Computational Biology **4**, e1000119 (2008).

[40] K. W. Latimer, D. Barbera, M. Sokoletsky, B. Awwad, Y. Katz, I. Nelken, I. Lampl, A. L. Fairhall, and N. J. Priebe, Multiple timescales account for adaptive responses across sensory cortices, Journal of Neuroscience **39**, 10019 (2019).

[41] R. A. Mease, M. Famulare, J. Gjorgjieva, W. J. Moody, and A. L. Fairhall, Emergence of adaptive computation by single neurons in the developing cortex, Journal of Neuroscience **33**, 12154 (2013).

[42] M. Díaz-Quesada and M. Maravall, Intrinsic mechanisms for adaptive gain rescaling in barrel cortex, Journal of Neuroscience **28**, 696 (2008).

[43] K. S. Gaudry and P. Reinagel, Benefits of contrast normalization demonstrated in neurons and model cells, Journal of Neuroscience **27**, 8071 (2007).

[44] C. A. Matulis, J. Chen, A. D. Gonzalez-Suarez, R. Behnia, and D. A. Clark, Heterogeneous temporal contrast adaptation in drosophila direction-selective circuits, Current Biology **30**, 222 (2020).

[45] K. J. Kim and F. Rieke, Temporal contrast adaptation in the input and output signals of salamander retinal ganglion cells, Journal of Neuroscience **21**, 287 (2001).

[46] Y. Yu and T. S. Lee, Dynamical mechanisms underlying contrast gain control in single neurons, Physical Review E **68**, 011901 (2003).

[47] C. Morris and H. Lecar, Voltage oscillations in the barnacle giant muscle fiber, Biophysical journal **35**, 193 (1981).

[48] K. S. Gaudry and P. Reinagel, Contrast adaptation in a nonadapting lgn model, Journal of neurophysiology **98**, 1287 (2007).

[49] E. M. Izhikevich, *Dynamical systems in neuroscience* (MIT press, 2007).

[50] See Supplemental Material at http://link.aps.org/supplemental/ for Secs. 1, 2, and Figs. S1-S7.

[51] A. L. Fairhall, G. D. Lewen, W. Bialek, and R. R. de Ruyter van Steveninck, Efficiency and ambiguity in an adaptive neural code, Nature **412**, 787 (2001).

[52] C. D. Wilson, G. O. Serrano, A. A. Koulakov, and D. Rinberg, A primacy code for odor identity, Nature communications **8**, 1477 (2017).

[53] R. Shusterman, M. C. Smear, A. A. Koulakov, and D. Rinberg, Precise olfactory responses tile the sniff cycle, Nature neuroscience **14**, 1039 (2011).

[54] J. Clemens, N. Ozeri-Engelhard, and M. Murthy, Fast intensity adaptation enhances the encoding of sound in drosophila, Nature communications **9**, 134 (2018).

[55] J. Benda, Neural adaptation, Current Biology **31**, R110 (2021).

[56] K. I. Nagel and A. J. Doupe, Temporal processing and adaptation in the songbird auditory forebrain, Neuron **51**, 845 (2006).

[57] D. H. Foster, Color constancy, Vision research **51**, 674 (2011).

[58] T. Kurahashi and A. Menini, Mechanism of odorant adaptation in the olfactory receptor cell, Nature **385**, 725 (1997).

[59] S. M. Block, J. E. Segall, and H. C. Berg, Impulse responses in bacterial chemotaxis, Cell **31**, 215 (1982).

[60] T.-M. Yi, Y. Huang, M. I. Simon, and J. Doyle, Robust perfect adaptation in bacterial chemotaxis through integral feedback control, Proceedings of the National Academy of Sciences **97**, 4649 (2000).

[61] A. I. Weber, K. Krishnamurthy, and A. L. Fairhall, Coding principles in adaptation, Annual review of vision science **5**, 427 (2019).

[62] H. Guo and D. P. Smith, Odorant receptor desensitization in insects, Journal of Experimental Neuroscience **11**, 1179069517748600 (2017).

[63] D. Wicher and F. Miazzi, Functional properties of insect olfactory receptors: ionotropic receptors and odorant receptors, Cell and Tissue Research **383**, 7 (2021).

[64] J. Benda and A. V. Herz, A universal model for spike-frequency adaptation, Neural computation **15**, 2523 (2003).

[65] L. Mukunda, F. Miazzi, S. Kaltofen, B. S. Hansson, and D. Wicher, Calmodulin modulates insect odorant receptor function, Cell calcium **55**, 191 (2014).

[66] V. Sargsyan, M. N. Getahun, S. L. Llanos, S. B. Olsson, B. S. Hansson, and D. Wicher, Phosphorylation via pkc regulates the function of the drosophila odorant co-receptor, Frontiers in cellular neuroscience **5**, 5 (2011).

[67] N. Kadakia and T. Emonet, Front-end weber-fechner gain control enhances the fidelity of combinatorial odor coding, Elife **8**, e45293 (2019).

[68] A. Egea-Weiss, C. J. Kleineidam, P. Szyszka, *et al.*, High precision of spike timing across olfactory receptor neurons allows rapid odor coding in drosophila, IScience **4**, 76 (2018).

[69] A. Hudspeth, F. Jülicher, and P. Martin, A critique of the critical cochlea: Hopf—a bifurcation—is better than none, Journal of neurophysiology **104**, 1219 (2010).

[70] A. Erez, T. A. Byrd, M. Vennettilli, and A. Mugler, Cell-to-cell information at a feedback-induced bifurcation point, Physical review letters **125**, 048103 (2020).

[71] B. Ermentrout, Linearization of fi curves by adaptation, Neural computation **10**, 1721 (1998).

[72] A. Rauch, G. La Camera, H.-R. Luscher, W. Senn, and S. Fusi, Neocortical pyramidal cells respond as integrate-and-fire neurons to in vivo–like input currents, Journal of neurophysiology **90**, 1598 (2003).

[73] D. G. Albrecht and D. B. Hamilton, Striate cortex of monkey and cat: contrast response function., Journal of neurophysiology **48**, 217 (1982).

[74] B. Wark, B. N. Lundstrom, and A. Fairhall, Sensory adaptation, Current opinion in neurobiology **17**, 423 (2007).

[75] N. Brenner, W. Bialek, and R. d. R. Van Steveninck, Adaptive rescaling maximizes information transmission, Neuron **26**, 695 (2000).

[76] J. L. Gardner, P. Sun, R. A. Waggoner, K. Ueno, K. Tanaka, and K. Cheng, Contrast adaptation and representation in human early visual cortex, Neuron **47**, 607 (2005).

[77] D. Chander and E. Chichilnisky, Adaptation to temporal contrast in primate and salamander retina, Journal of Neuroscience **21**, 9904 (2001).

[78] F. Rieke, Temporal contrast adaptation in salamander bipolar cells, Journal of neuroscience **21**, 9445 (2001).

[79] I. R. Graf and B. B. Machta, A bifurcation integrates information from many noisy ion channels and allows for milli-kelvin thermal sensitivity in the snake pit organ, Proceedings of the national academy of sciences **121**, e2308215121 (2024).

[80] J. M. Keegstra, F. Avgidis, Y. Mulla, J. S. Parkinson, and T. S. Shimizu, Near-critical tuning of cooperativity revealed by spontaneous switching in a protein signalling array, bioRxiv , 2022 (2022).

[81] D. M. Sherry, I. R. Graf, S. J. Bryant, T. Emonet, and B. B. Machta, Lattice ultrasensitivity produces large gain in e. coli chemosensing, arXiv preprint arXiv:2405.18331 (2024).

[82] A. Hudspeth, Integrating the active process of hair cells with cochlear function, Nature Reviews Neuroscience **15**, 600 (2014).

[83] S. Camalet, T. Duke, F. Jülicher, and J. Prost, Auditory sensitivity provided by self-tuned critical oscillations of hair cells, Proceedings of the national academy of sciences **97**, 3183 (2000).

[84] Y. Choe, M. O. Magnasco, and A. Hudspeth, A model for amplification of hair-bundle motion by cyclical binding of ca2+ to mechanoelectrical-transduction channels, Proceedings of the National Academy of Sciences **95**, 15321 (1998).

[85] V. M. Eguíluz, M. Ospeck, Y. Choe, A. J. Hudspeth, and M. O. Magnasco, Essential nonlinearities in hearing, Physical review letters **84**, 5232 (2000).

[86] A. S. Momi, M. C. Abbott, J. Rubinfien, B. B. Machta, and I. R. Graf, Hair cells in the cochlea must tune resonant modes to the edge of instability without destabilizing collective modes, bioRxiv , 2024 (2024).

[87] C. Briat, A. Gupta, and M. Khammash, Antithetic integral feedback ensures robust perfect adaptation in noisy biomolecular networks, Cell systems **2**, 15 (2016).

[88] R. T. Cardé and M. A. Willis, Navigational strategies used by insects to find distant, wind-borne sources of odor, Journal of chemical ecology **34**, 854 (2008).

[89] S. D. Stupski and F. van Breugel, Wind gates search states in free flight, bioRxiv , 2024 (2024).

[90] G. La Camera, A. Rauch, W. Senn, H.-R. Lüscher, and S. Fusi, Firing rate adaptation without losing sensitivity to input fluctuations, in *Artificial Neural Networks—ICANN 2002: International Conference Madrid, Spain, August 28–30, 2002 Proceedings 12* (Springer, 2002) pp. 180–185.

[91] https://github.com/emonetlab/bifurcation-temporal-information.

[92] B. Ermentrout, *Simulating, Analyzing, and Animating Dynamical Systems: A Guide to XPPAUT for Researchers and Students*, Software, environments, tools (Society for Industrial and Applied Mathematics, Philadelphia, 2002) pp. xiv, 290.

# Supplemental Materials for: Bifurcation enhances temporal information encoding in the olfactory periphery

Kiri Choi,[1, 2, 3, *] Will Rosenbluth,[1, *] Isabella R. Graf,[2, 4, 5] Nirag Kadakia,[1, 2, 3, †] and Thierry Emonet[1, 2, 4, 6, ‡]

[1]*Department of Molecular, Cellular, and Developmental Biology,*
*Yale University, New Haven, Connecticut 06511, USA*
[2]*Quantitative Biology Institute, Yale University, New Haven, Connecticut 06511, USA*
[3]*Swartz Foundation for Theoretical Neuroscience,*
*Yale University, New Haven, Connecticut 06511, USA*
[4]*Department of Physics, Yale University, New Haven, Connecticut 06511, USA*
[5]*Developmental Biology Unit, European Molecular Biology Laboratory, 69117 Heidelberg, Germany*
[6]*Interdepartmental Neuroscience Program, Yale University, New Haven, Connecticut 06511, USA*

## I. MI IS NOT SENSITIVE TO BINNING IN THE CONTEXT OF OUR ANALYSIS

Numerical calculation of MI is prone to the binning process. Either too small or too large bin sizes can introduce errors to the estimated MI, especially when the number of data points is finite and contains missing values (e.g., not all values in one variable have a corresponding pair in another variable). Fortunately, we can generate signals for an arbitrarily long time and simulate the respective firing rate perfectly. We find that our system operates in a regime where the histograms of the variables are smooth, such that the histogram-based numerical MI calculation is robust enough.

According to Steuer et al. [1], the systematic error introduced in MI by the binning process can be estimated by

$$err = \frac{m_x m_y - m_x - m_y + 1}{2N}. \quad (1)$$

Here, $m$ denotes the number of discrete states for two variables $(x, y)$, while $N$ denotes the number of data points. In our study, with $m_{stim} = m_{rate} = 100$, $m_{stim} \cdot m_{rate} = 10^4$, and $N \geq 10^6$, our estimated error in MI is around 0.0049, which is significantly smaller than the scale of MI.

Additionally, we scanned through combinations of bin numbers $m_{stim}$ and $m_{rate}$ to ensure that our system is operating in a smooth regime. We find that for all combinations of bin numbers, MI stays consistent: MI goes over 1 bit and responds similarly to the increase in signal variance (Fig. S1). The combinations of bin numbers tested include those suggested by various well-established systematic criteria [2].

## II. THE REVISED LINEAR-NONLINEAR MODEL EXHIBITS ELEVATED CODING CAPACITY AND VARIANCE ADAPTATION

Linear-nonlinear models are widely used to describe the phenomenological properties of neural dynamics and were previously used to emulate signal variance adaptation [3]. A traditional LN model includes two parts: a convolution with a smoothing filter (e.g., Gaussian), followed by a nonlinear compression (e.g., a rectified linear unit, ReLU), which simulates the effect of saturation [Fig. S4(a), red]. Under this framework, the model does not exhibit variance adaptation [Fig. S4(b), red]. However, when we substitute the rectified linear unit for the SNIC bifurcating $I$-$r$ curve [Fig. 1(b)], the imperfect gain control is recovered [Fig. S4(b), orange]. Once the input $I$ crosses the spiking threshold, $dr/dI$ decreases as $I$ increases, pushing the system farther away from the bifurcation point, such that greater fluctuations in $I$ are required to drive the same change in firing rate. In Hodgkin-Huxley type neurons, the time to spike reflects the weighted average of stimulus intensity over the relatively small inter-spike intervals. However, to construct a continuous firing rate from discrete firing events, the spiking events must be smoothed by a filter whose width is significantly greater than the inter-spike period. In terms of LN models, this means applying linear filtering after the nonlinear compression to make our analyses analogous to that of the Hodgkin-Huxley type model. This new architecture, which we call the nonlinear-linear (NL) model [Fig. S4(a), green and blue], retains variance adaptation [Fig. S4(b), green and blue] and demonstrates a significant improvement ($\sim$ 15%) in coding capacity with respect to their LN model counterparts [Fig. S4(c)]. Replacing the architecture from LN to NL smoothens the transition from activity to quiescence, allowing information about the time since the excursion into the quiescent regime to be extracted. With a correlated input like an O-U signal, the system can encode information about the stimulus even when the neuron is close to the quiescent regime, providing distinct advantages over traditional LN models.

———————
* These authors contributed equally to this work
† nirag.kadakia@yale.edu
‡ thierry.emonet@yale.edu

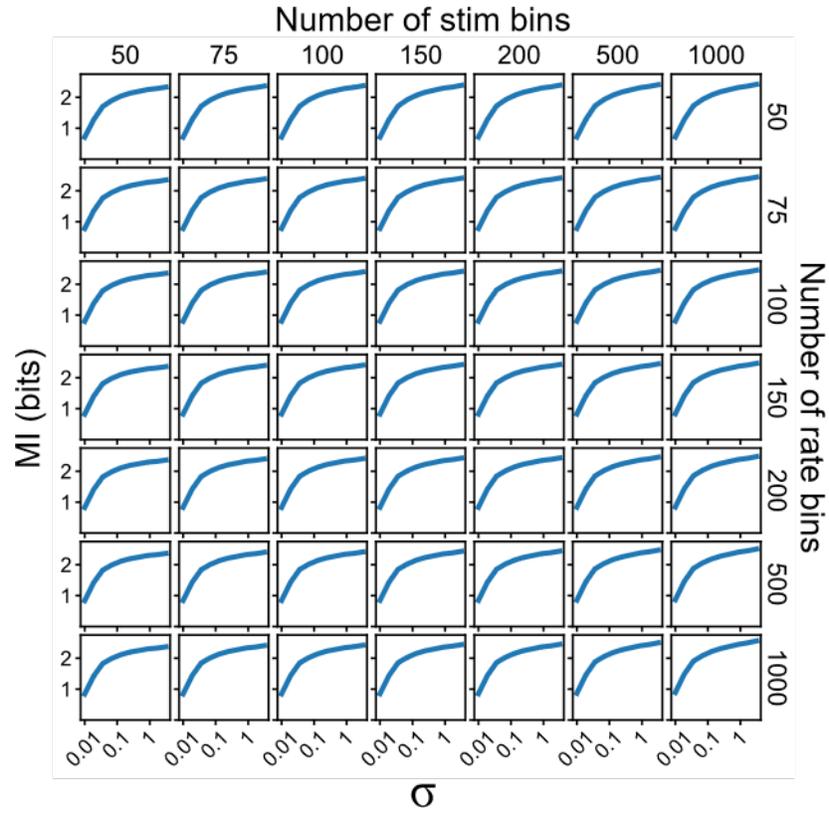

Fig. S1. MI encodes more than 1 bit of information regardless of the bin size. Discretized MI calculation can be influenced by the number of bins, but for our system, MI seems to stay relatively consistent. $I_b = 4.54 pA$ is used.

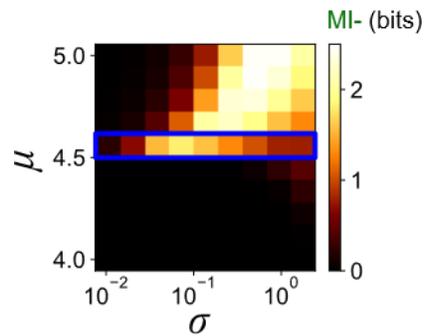

Fig. S2. The variance adaptation is due to the proximity to a bifurcation and not the signal process. MI- as a function of the mean $\mu$ and standard deviation $\sigma$ using a Gaussian white noise as the input current. The blue box corresponds to $\mu \sim I_b = 4.54 pA$. When $\mu$ is close to $I_b = 4.54 pA$, MI- stays elevated over the wide range of $\sigma$.

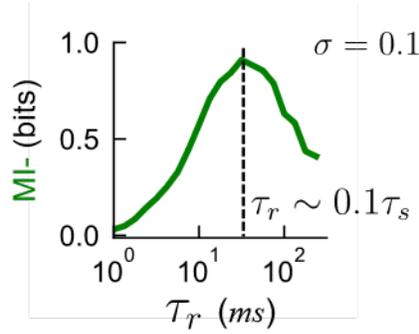

Fig. S3. MI- is maximized around $\tau_r \sim 1/10\tau_s$. A vertical cross-section of Fig. 3(c) at $\sigma \sim 0.1$. This relationship is consistent across different $\sigma$.

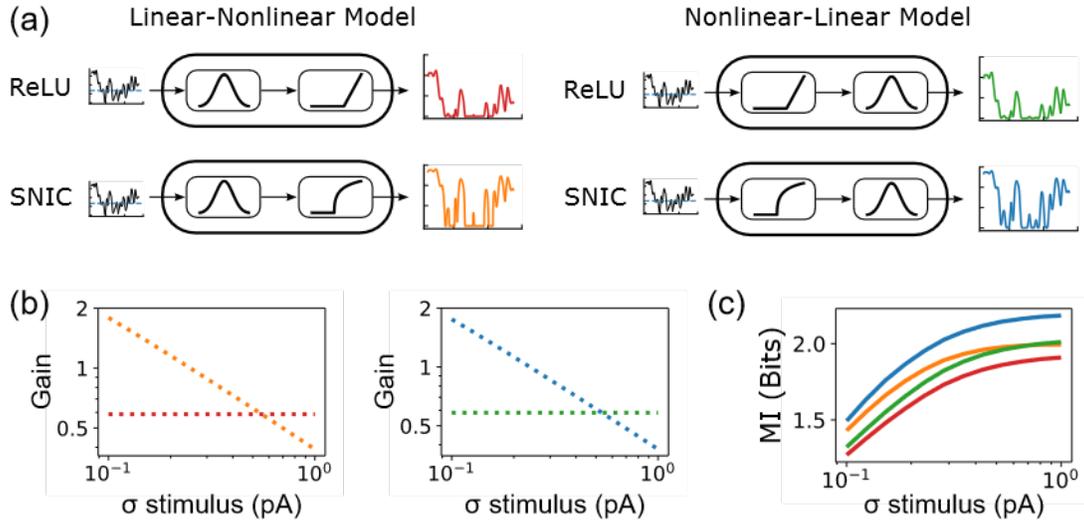

Fig. S4. Using a SNIC bifurcation in a nonlinear-linear architecture model allows variance adaptation and high information throughput. (a) Four different models can be generated using a combination of ReLU or SNIC nonlinearities in LN or NL architecture. Red: LN-ReLU; orange: LN-SNIC; green: NL-ReLU; blue: NL-SNIC. (b) Gain, defined as the ratio of the standard deviation of the model response to the standard deviation of the stimulus, plotted for LN models and NL models for an O-U stimulus with $\sigma$ ranging from 0.1 to 1 $pA$. Although the order of operations of the model architecture does not affect the gain, gain control is only observed in models that use the SNIC filter (orange and blue). (c) The mutual information between the stimulus and the response for the four models. The same color code is used. NL model with the SNIC filter encodes the highest amount of information.

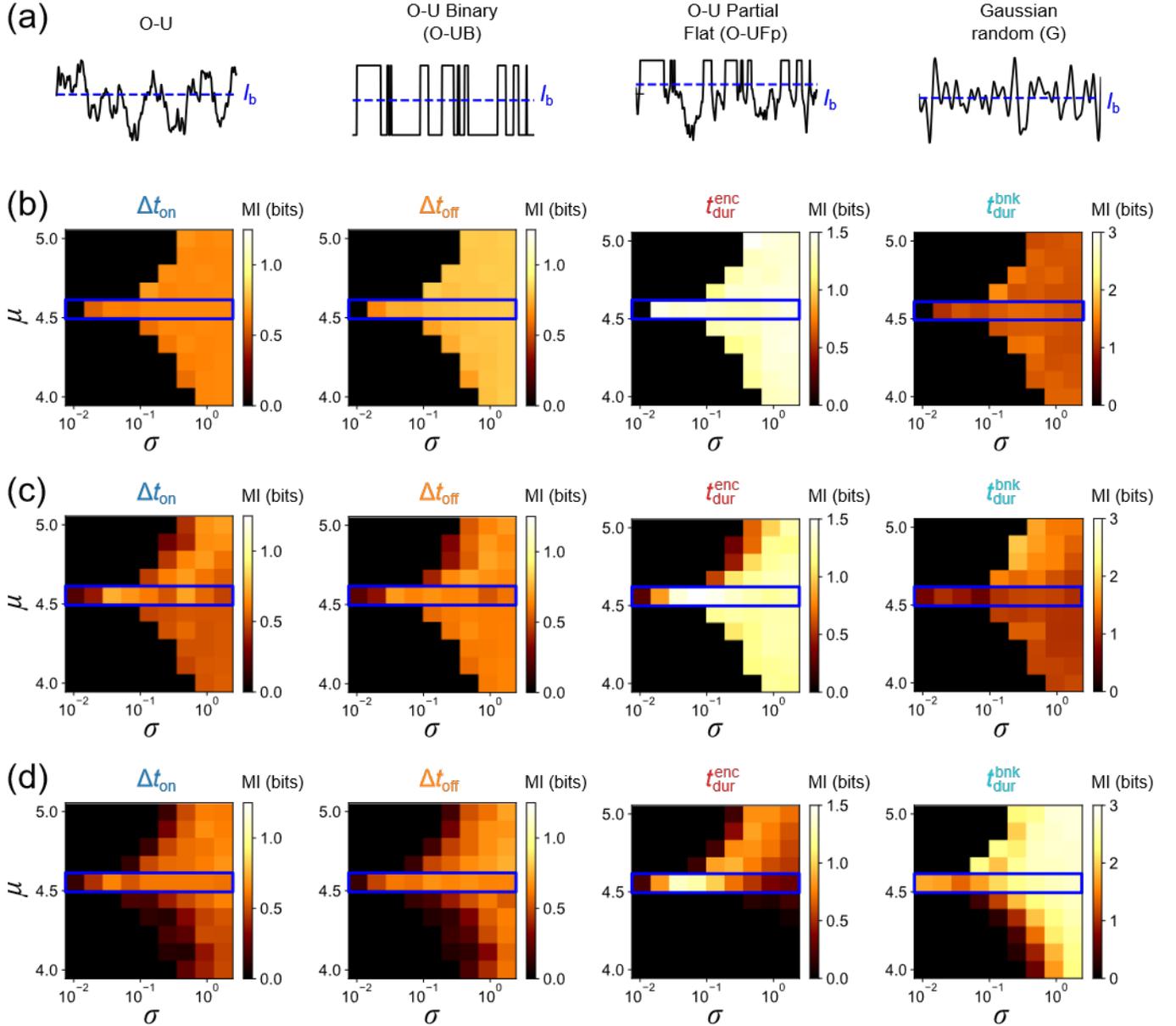

Fig. S5. The coding capacity for signal timing is influenced by the statistical properties of signal crossings and does not depend on a specific signal-generating process. (a) A schematic illustrating traces of different signal types, including the Ornstein–Uhlenbeck process (O-U), a binarized Ornstein–Uhlenbeck process (O-UB), a partially flattened Ornstein–Uhlenbeck process (O-UFp), and a Gaussian random (G). (b) MI between the distributions of encounter times $\Delta t_{ON}$ (first column), blank times $\Delta t_{OFF}$ (second column), encounter duration $t_{dur}^{enc}$ (third column), and blank duration $t_{dur}^{bnk}$ (fourth column) and the firing rate $r$ as a function of signal mean $\mu$ and standard deviation $\sigma$ using a fully-binarized O-UB input. (c) Same as (b) for a partially reduced O-UFp as the input signal, where only the fluctuations above $I_b$ are removed. (d) Same as (b) for Gaussian white noise as the input signal. Note the changes in color scales for different temporal statistics.

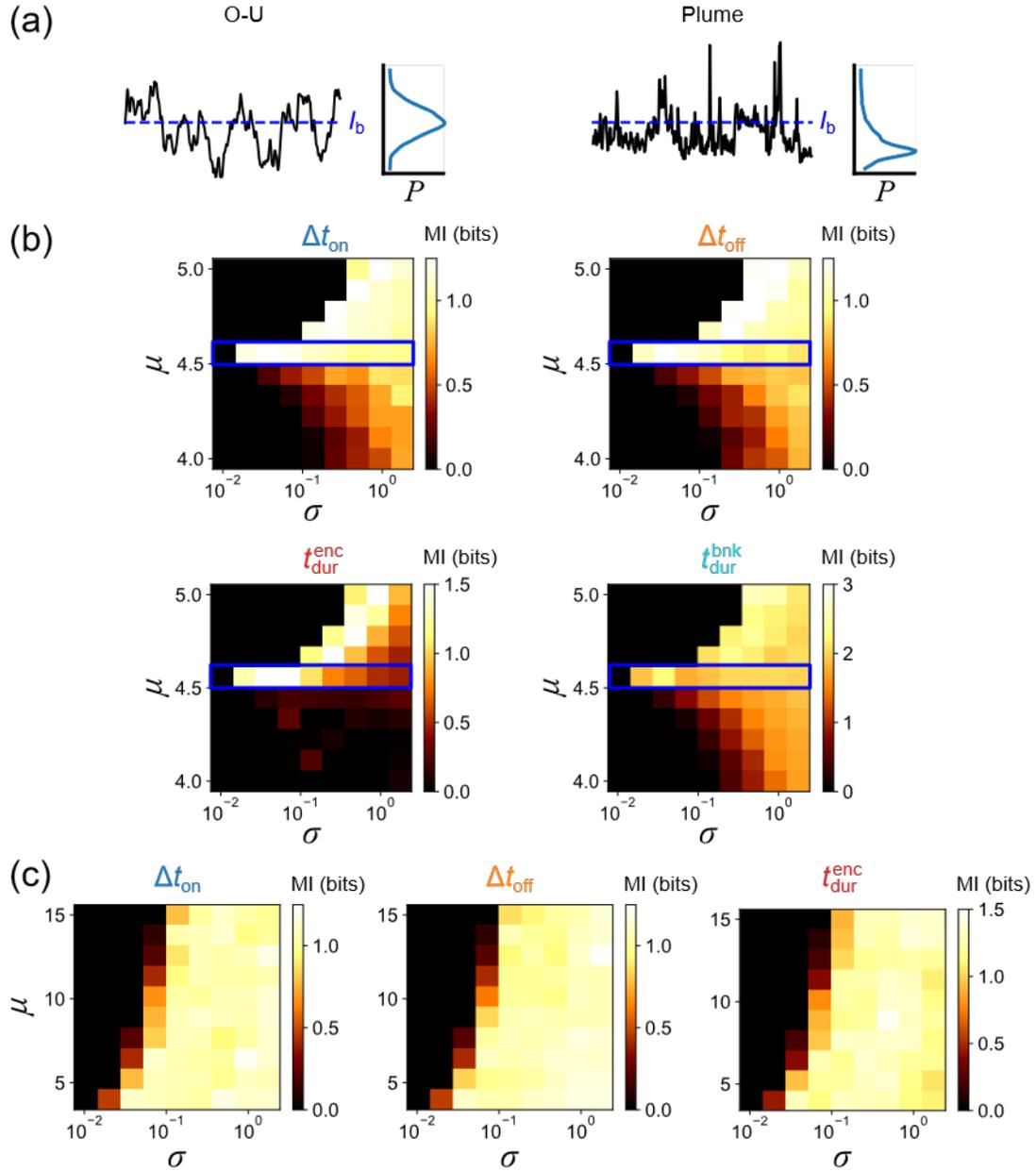

Fig. S6. The coding capacity for signal timing using experimentally measured odor traces encountered by freely navigating flies. (a) The odor signal fruit flies experience in turbulent plumes differs from the O-U signal. The time trace of real plume concentrations encountered by an animal shows a long-tailed distribution, and the distributions of odor encounter and blank durations exhibit near power-law behavior. (b) MI between the distributions of encounter times $\Delta t_{\text{ON}}$, blank times $\Delta t_{\text{OFF}}$, encounter duration $t_{\text{dur}}^{\text{enc}}$, and blank duration $t_{\text{dur}}^{\text{bnk}}$ and the firing rate $r$ as a function of signal mean $\mu$ and standard deviation $\sigma$ using a real plume signal in the non-adapting Na+K model. The real plume time traces are shifted and rescaled to fit the denoted $\mu$ and $\sigma$. (c) Same as (b) but using the adapting Na+K+Ca model with the firing rate threshold of $\sim 30\ Hz$. Note the changes in color scales for different temporal statistics.

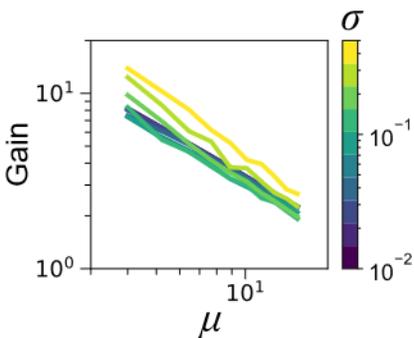

Fig. S7. The biophysical model exhibits gain scaling consistent with the Weber-Fechner law. The color bar corresponds to the $\sigma$ from 0.01 to 0.5.

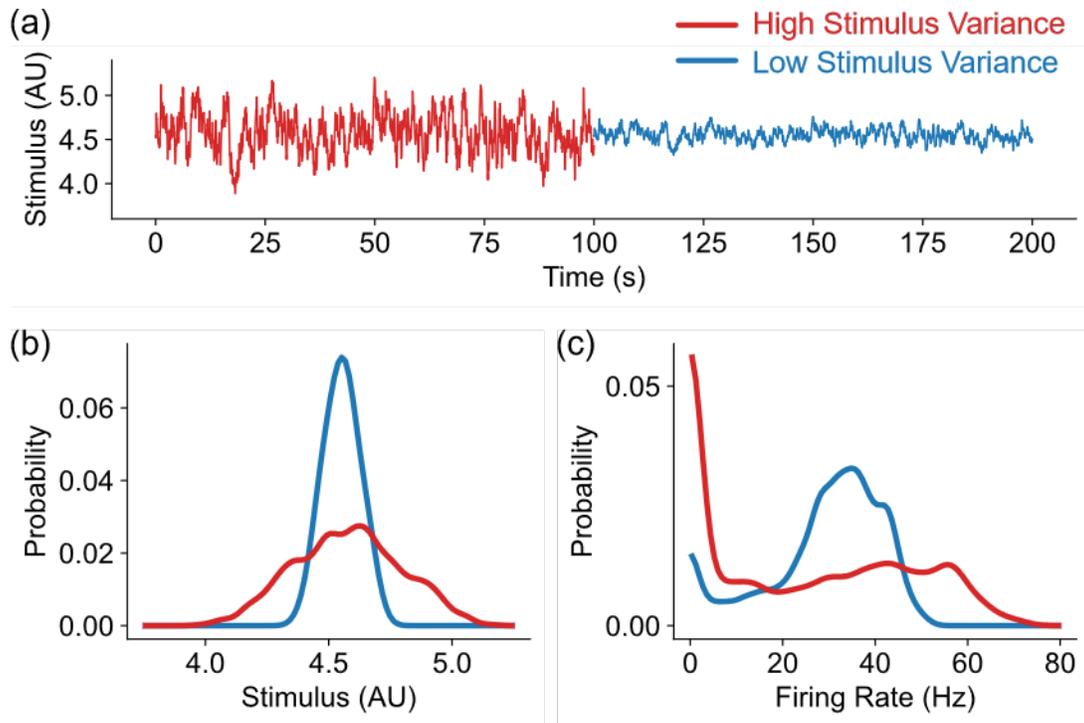

Fig. S8. The response of the biophysical model to changes in the signal variance qualitatively reproduces the experimental measurements from [4] Figure 4a,b,d. (a) Simulated time trace of the odor signal with high variance ($\sigma = 0.24$) and low variance ($\sigma = 0.08$). The signal mean ($\mu = 4.54$) is kept the same. (b) The distributions of the simulated odor signal with high variance (red) and low variance (blue). (c) The corresponding distributions of the firing rate simulated using the biophysical model.

[1] R. Steuer, J. Kurths, C. O. Daub, J. Weise, and J. Selbig, The mutual information: detecting and evaluating dependencies between variables, Bioinformatics **18**, S231 (2002).

[2] C. J. Cellucci, A. M. Albano, and P. E. Rapp, Statistical validation of mutual information calculations: Comparison of alternative numerical algorithms, Physical review E **71**, 066208 (2005).

[3] A. I. Weber, K. Krishnamurthy, and A. L. Fairhall, Coding principles in adaptation, Annual review of vision science **5**, 427 (2019).

[4] S. Gorur-Shandilya, M. Demir, J. Long, D. A. Clark, and T. Emonet, Olfactory receptor neurons use gain control and complementary kinetics to encode intermittent odorant stimuli, Elife **6**, e27670 (2017).